\documentclass[a4paper,11pt]{article}
\usepackage{pos}
\usepackage{subcaption}

\title{Run 2/3 measurement of the muon anomalous magnetic moment by the Muon g-2 experiment at Fermilab}

\author*[a]{Estifa'a Zaid}

\affiliation[a]{University of Liverpool,\\
 Oliver Lodge, Liverpool, United Kingdom}

\affiliation[*]{On behalf of the g-2 collaboration}

\emailAdd{ezaid@liverpool.ac.uk}

\abstract{The Muon g-2 experiment at Fermilab seeks to measure the muon magnetic moment anomaly, $a_\mu =(g-2)/2$, with a final target precision of 0.14 parts-per-million (ppm). The experiment’s initial result,
published in 2021 using Run 1 data from 2018, confirmed the previous measurement at Brookhaven
National Laboratory with a comparable sensitivity of 0.46 ppm. In 2023, a new result based on Run 2 and Run 3 data, collected in 2019 and 2020, was released. These datasets contain four times the
data from Run 1, significantly enhancing sensitivity and achieving an unprecedented uncertainty of
0.20 ppm. This advancement resulted in a two-fold improvement in both statistical and systematic
uncertainties. Here, we will discuss the muon $g-2$ measurement, the increased precision relative to the Run 1 result, and provide an outlook on future measurements which will incorporate datasets from 2021 to 2023. Additionally, we will explore the implications of comparing the new measurements with the latest Standard Model predictions for muon g-2.}

\FullConference{The 43rd International Symposium on Physics in Collision (PIC2024)\\
 22–25 October 2024\\
Athens, Greece\\}

\begin{document}
\maketitle

\section{The Muon g-2}

The magnetic moment $\vec{\mu}$ of a fundamental particle with mass $m$, charge $q$ and spin $\vec{S}$ can be expressed 

\begin{equation}
    \vec{\mu} = g\frac{q}{2m}\vec{S}
\end{equation}
where $g$ is the gyromagnetic ratio, a dimensionless quantity which describes the strength of the magnetic moment. Schwinger’s calculation of the electron gyromagnetic ratio in 1948 demonstrated that $g$ differs from the Dirac theory prediction of 2. \cite{Dirac1928}. This difference is due to the interaction of a given lepton with virtual particles which provide an additional contribution to the g-factor which can be parametrised in terms of the anomalous magnetic moment $a = \frac{g-2}{2}$. 

The electron g-factor has been calculated to an astonishing precision, and the electron anomalous magnetic moment is found to agree with the experimental measurement of its value \cite{Fan2023}. The muon anomalous magnetic moment however has provided physicists with a puzzle that has spanned over 20 years. A discrepancy between the experimental measurement and Standard Model (SM) prediction persists, indicating the possible presence of particles not included in the SM theory.  

The E821 experiment conducted at Brookhaven National Laboratory (BNL) \cite{Bennett2006} measured the muon magnetic anomaly with a relative precision of 0.54 parts per million (ppm), revealing a discrepancy of nearly $3 \sigma$ from the SM prediction \cite{Venanzoni2023}. Over time, improvements in the SM calculations increased this discrepancy to $3.7 \sigma$, marking it as one of the most pronounced differences between experimental data and SM predictions. To investigate this discrepancy further and improve the precision of the measurement, the E989 muon g-2 experiment was launched at Fermi National Accelerator Laboratory  (FNAL), beginning data collection in 2018. The two published results from the Fermilab experiment \cite{Abi2021} \cite{PhysRevLett2023} will be discussed here together with improvements expected for the third and final published result expected in 2025. The high precision achieved by this experiment has motivated the theory community to produce ever more precise SM predictions for muon g-2.

\section{The Muon g-2 Experiment at Fermilab}

The Muon g-2 experiment at Fermilab was set up with the design aim of measuring the anomalous magnetic moment of the muon to a precision of 0.14 ppm, a fourfold improvement in precision with respect to the BNL experiment \cite{grange2018}. The experiment uses positively charged muons produced by the Fermilab accelerator complex. Fermilab houses a linear accelerator where protons reach energies up to 400 MeV before being injected into a booster ring where they can reach 8 GeV. A fixed target is used to produce high energy positive pions; 3.1 GeV pions are then selected and separated from protons. The resulting muon beam, a product of pion decay,  is 95\% polarised, a feature that is crucial to the experimental principle of the muon g-2 experiment.  

\subsection{Experimental Setup}

\begin{figure}
    \centering
    \includegraphics[width=0.65\linewidth]{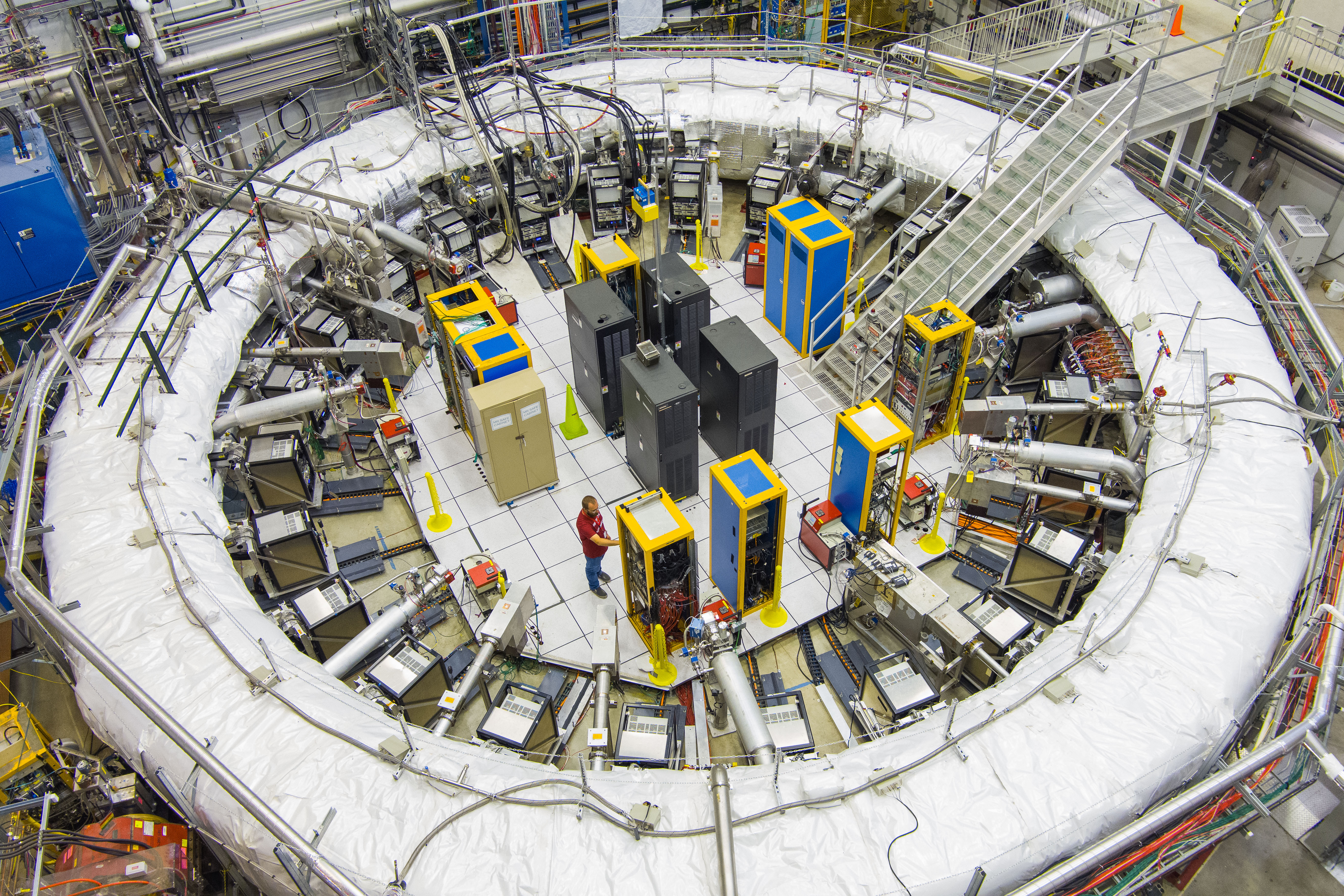}
    \caption{View of the muon g-2 experimental setup at Fermilab. The most notable feature is the storage ring covered in a white insulating blanket \cite{Venanzoni2023}.}
    \label{fig:ring}
\end{figure}

The main feature of the muon g-2 experimental apparatus is a 3.56 m radius 1.45 T superferric magnetic storage ring which has been recycled from the BNL experiment. The ring is used to confine the polarised muon beam through a series of three superconducting coils that span the circumference of the steel yoke providing a uniform magnetic field throughout. Figure \ref{fig:ring} shows an aerial view of this ring with insulation in place. The muons are confined vertically using four electrostatic quadrupoles that are placed symmetrically around the ring. 

There are three types of detectors used to make this precision measurement. The first type consists of 24 electromagnetic calorimeters made of lead fluoride ($\mathrm{PbF_2}$) crystals spaced around the interior of the ring \cite{Khaw2019}; these are used to measure the energy of the positrons decaying from the muons. Each crystal is coupled to a silicon photomultiplier readout. The second type consists of two straw tube tracker stations which are used to measure properties of the muon beam in order to better determine systematic uncertainties and perform auxiliary measurements of the muon beam. Each tracking station consists of 8 tracking modules which all sit within the storage ring vacuum but outside of the muon beam path. The third type consists of Nuclear Magnetic Resonance (NMR) probes which are used to measure the magnetic field. There are 378 probes distributed across 72 location above and below the storage ring which continuously measure field drifts during muon storage. A so-called "field trolley" is pulled through the storage ring every 3 to 5 days when there is no muon beam. This provides a map of the field and consists of 17 NMR probes. In addition to the detector setups outlined, a state-of-the-art laser calibration system is used to measure the gain fluctuations at sub-per-mill level precision over a range of timescales from $10 ns$ to the lifetime of the whole experiment. This is done by sending simultaneous calibration pulses onto each of the 1296 crystals of the electromagnetic calorimeter. 

\subsection{Experimental principle}

A charged particle, when placed in a magnetic field $\vec{B}$, will traverse in a circular orbit, and its spin will feel a torque causing it to act like a spinning top. As a muon traverses the ring its spin direction precesses ahead of its momentum direction which rotates at the cyclotron frequency. The experiment measures the anomalous precession frequency defined as the difference between the two; this frequency, together with the magnetic field,  allows us to determine $a_{\mu}$

\begin{equation}
    \omega_{a} = \omega_{s}-\omega_{c} = a_{\mu}\frac{e}{m_{\mu}c}B.
\end{equation}
The frequency of precession $\omega_{s}$ and the strength of the magnetic field provide the components to calculate the anomalous magnetic moment $a_{\mu}$. In such an experiment where we determine $g$ by measuring $a_{\mu}$ directly; if g were equal to 2 then $a_{\mu}=0$ and the spin precession would be equal to the cyclotron  frequency.

\begin{figure}
    \centering
    \includegraphics[width=0.7\linewidth]{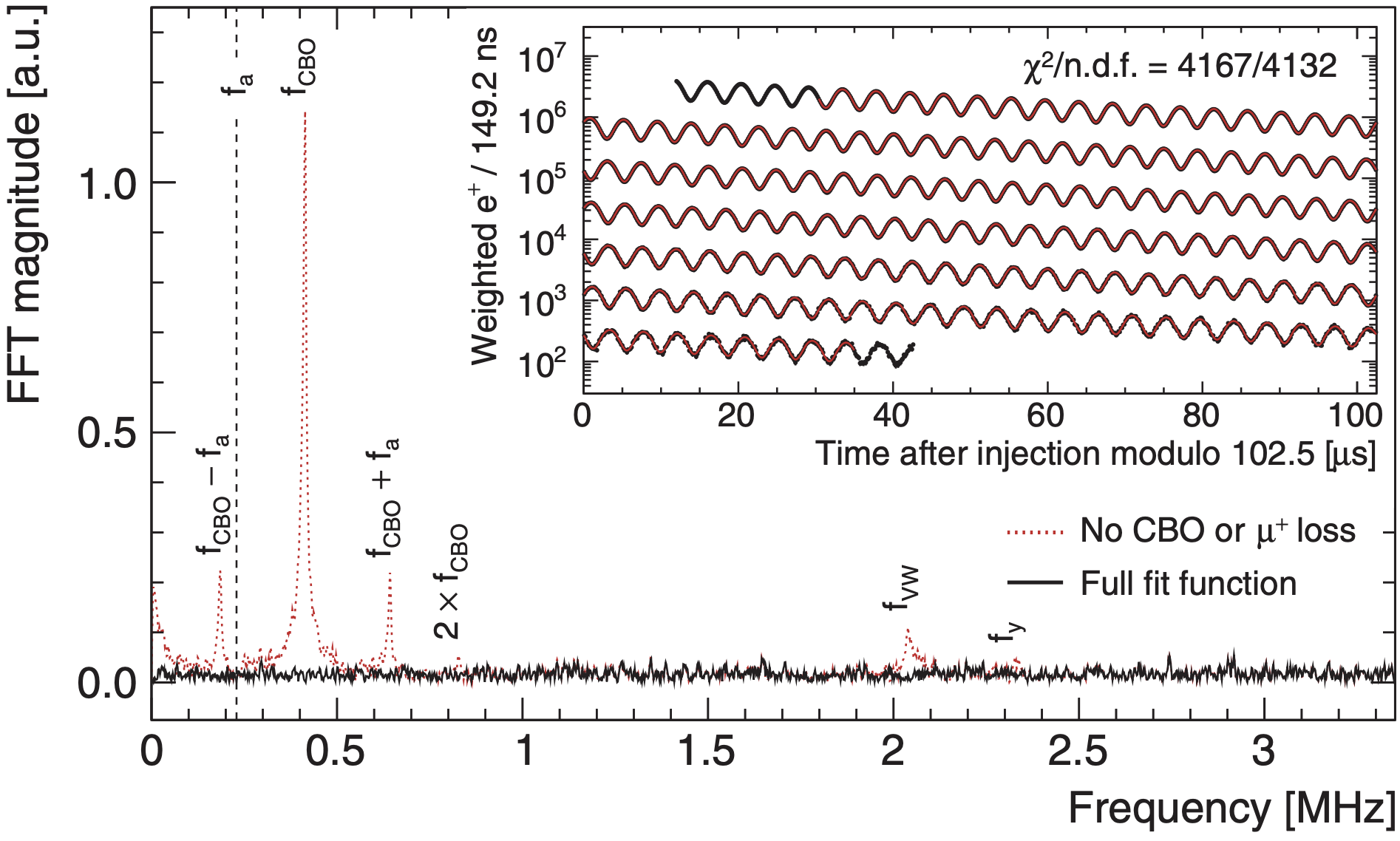}
    \caption{Fourier transform of the residuals from a time-series fit. The red dashed curve shows the result when betatron motion and muon losses are neglected, while the black curve corresponds to the full fit, including these effects. The inset shows $e^{+}$ time spectrum (black) from the Run 1c dataset, overlaid with the full fit function (red) \cite{Abi2021}.}
    \label{fig:wiggleFFT}
\end{figure}

Due to the spin polarisation of muons and parity violation in weak decays, high-energy positrons are preferentially emitted along the direction of the muon spin. As the muon spin precesses in the magnetic field, the detection and counting of positrons with energies above a specified threshold offer a means to measure the anomalous precession frequency, $\omega_{a}$. The number of high-energy positrons entering the calorimeters is counted as a function of time and then weighted according to the decay asymmetry that is correlated to the positron energy. This time spectrum (wiggle plot) has an oscillation that is at the $\omega_{a}$ frequency. The highest number of positrons occurs when the spin vector aligns with the momentum vector. To extract $\omega_{a}$, a multi-parameter fit function is performed while accounting for beam oscillations muon losses and detector effects. Each beam-dynamic effect contributes an additional frequency component to the wiggle plot and therefore increases the number of parameters in the fit.   
Figure \ref{fig:wiggleFFT} shows the oscillation and fit of the detected $e^{+}$ time spectrum as well a Fast Fourier Transform (FFT) of the residuals from time-series fits. 

The NMR probe is used to measure the magnetic field as a precession frequency (Larmor frequency) $\omega_{p}$ \cite{Fei1997}. The ratio of the two frequencies $\omega_{a} / \omega_{p}$ together with reference measurements give the anomalous magnetic moment of the muon, $a_{\mu}$

\begin{equation}
    a_{\mu} =  \frac{\omega_{a}}{\omega_{p}} \frac{\mu_{p}}{\mu_{e}}\frac{m_{p}}{m_{e}}\frac{g_{e}}{2}
    \label{eq:amuref}
\end{equation}
where external  well-known constants contribute an overall uncertainty of $\sim 25$ parts-per-billion (ppb) \cite{Tiesinga2021}. From equation \ref{eq:amuref}, one can take into account the motion of the muon beam and transient magnetic fields present in the ring. 

\begin{equation}
a_\mu = 
\underbrace{
\left[ 
\frac{f_\text{clock} \cdot \omega_a^m \left( 1 + C_e + C_p + C_{pa} + C_{dd} + C_{ml} \right)}
{f_\text{calib} \cdot \langle \omega_p(\vec{r}) \times M(\vec{r}) \rangle \left( 1 + B_q + B_k \right)} 
\right]
}_{\mathcal{R}_\mu}
\frac{\mu_p(T_r)}{\mu_e(H)} 
\frac{\mu_e(H)}{\mu_e} 
\frac{m_\mu}{m_e} 
\frac{g_e}{2}
\label{eq:amuall}
\end{equation}
Such effects lead to equation \ref{eq:amuall} where $\omega_p$ is fractionally corrected by the quad and kicker transients, $B_q$ and $B_k$ respectively.  $B_q$ describes the vibrations of the ESQ plates that disturb the magnetic field and $B_k$  the residual eddy current from the kicker.  Similarly, five correction factors are applied to the measured $\omega_a$. The QED factor, $\mu_e(H)/\mu_e $ , represents the ratio of the electron’s magnetic moment in a hydrogen atom to that of a free electron in vacuum. The reference temperature, $T_r = 34.7^{\circ}C$, is the condition at which the shielded proton-to-electron magnetic moment ratio,$\mu_p(T_r)/\mu_e(H)$, is measured.  Additionally, the muon-to-electron mass ratio $m_\mu/m_e$ ,  has been determined with a precision of 22 ppb using muonium spectroscopy. All of the corrections,  $C_i$,  arise from effects that cause a frequency bias. These originate from the behaviour of the muon beam as well as the energy dependence of the positron drift time and the time dependence of calorimeter acceptance. $C_e$ is the electric field correction needed to account for the spread of the momenta of muons and arises from non-zero electric fields. $C_p$ is the pitch correction which accounts for the vertical oscillation of muons. $C_{ml}$ is the muon loss correction arising from the decay rate of the muons being momentum dependent therefore, affecting the muon initial phase over fill time and biasing $\omega_a$. $C_{pa}$ is the phase acceptance correction arising from the energy dependence of the positron drift time and the time dependence of the calorimeter acceptance. $C_{dd}$ is the  differential decay correction which is due to high-momentum muons having a longer lifetime.

The Muon g-2 collaboration employs a blinding technique to prevent unintentional biases in the analysis and to ensure the integrity of its results.  This is a hardware blinding done by introducing an unknown frequency offset ($\Delta f$) into the clock system used to measure the $\omega_a$ frequency. This offset is chosen by a separate independent team and is kept secret until all analysis steps are complete. The unblinding factor is denoted as $f_{clock}$.

Based on the measurement principle explained above, the analysis is divided into two parts; the measurement of the magnetic field strength $\omega_p$ and the measurement and corrections of the anomalous precession frequency $\omega_a$. To minimise errors and use diverse analysis techniques, the $\omega_a$ measurement is performed by several independent groups which are then combined at the end taking into account correlations between analysis methods.

\section{Run 1 result}

The Muon g-2 experiment at Fermilab took data for 6 years. Over that time, 322 billion positrons were collected and analysed, which corresponds to a statistical uncertainty that surpasses the experiment's design goal of 100 ppb statistical uncertainty and taking 21.9 times the amount of data as Brookhaven. This data can be categorised into 6 data-taking campaigns (Runs 1-6). The first experimental run began in March 2018 and results for the analysis of this data were published in April 2021 \cite{Abi2021}. 

The Run 1 analysis found $a_\mu$ to be $116592040 \pm 54$ $10^{-11}$, confirming the measurement made at Brookhaven to within $0.6\sigma$ and providing a greater tension of $4.2\sigma$ with the SM prediction at the time. Figure \ref{fig:syserrandresults} includes FNAL Run 1 result.  

The Run 1 analysis uncertainties were dominated by statistics, giving a statistical error of 460 ppb whilst the systematic uncertainty was 157 ppb. Both of these were significantly improved in the Run 2/3 analysis as will be explained in the subsequent section.

\section{Run 2/3 analysis improvements and results}

Run 2 and Run 3 data were taken in 2019-2020 and the Run 2/3 analysis result was published in August 2023. This measurement benefited from a greater number of statistics as well as substantial improvements to the systematic uncertainties. The statistical uncertainty was reduced from 460 ppb in Run 1  to 201 ppb in Run 2/3; this is directly due to 4.7 times more positron data being analysed. 

\begin{figure}[htbp]
	\centering
	\begin{subfigure}{0.48\linewidth}
		\includegraphics[width=\linewidth]{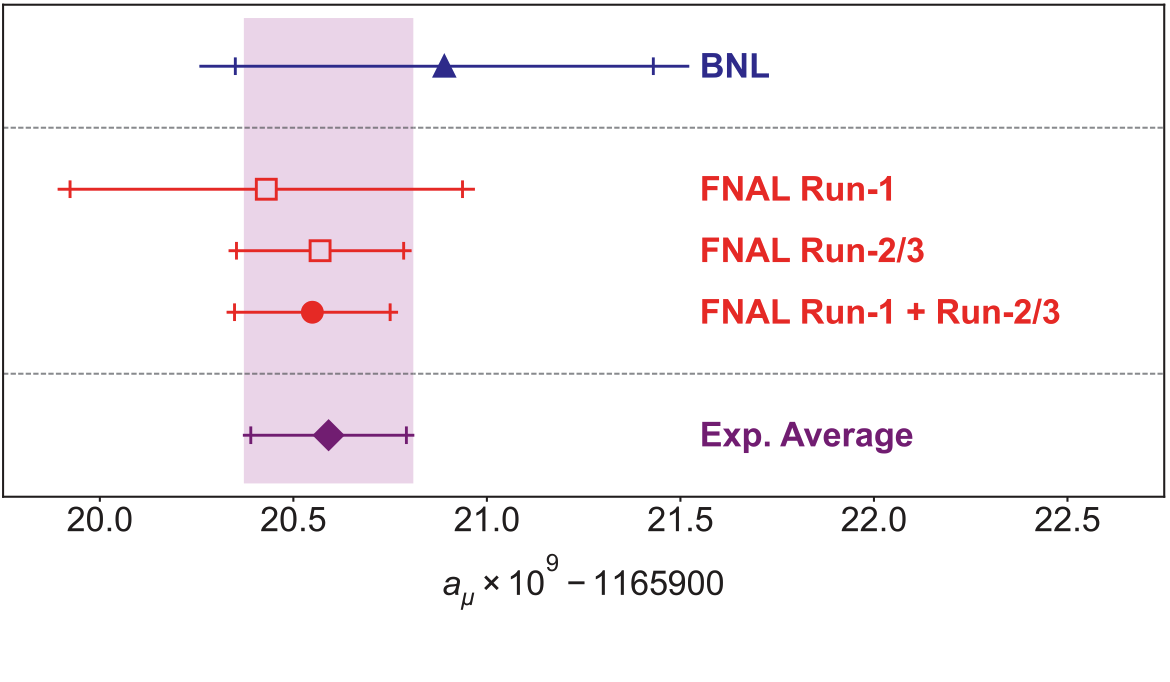}
        \label{fig:run1res}
	\end{subfigure}
	\begin{subfigure}{0.50\linewidth}
		\includegraphics[width=\linewidth]{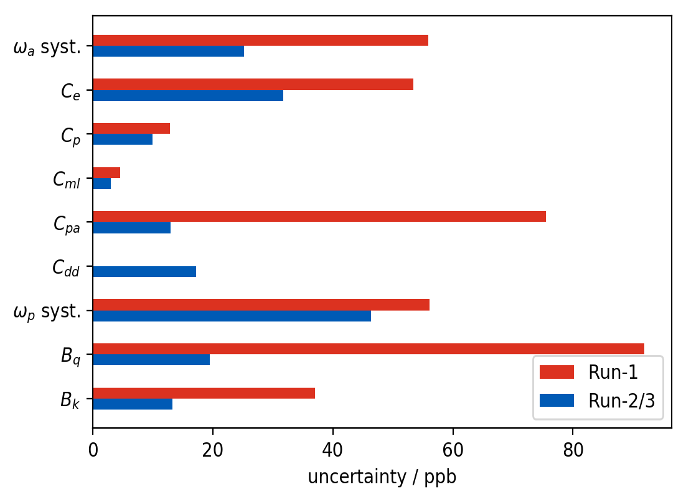}
        \label{fig:syserr}
	\end{subfigure}
    \caption{\textbf{Left},  Overview of the experimental results for $a_\mu$, the plot shows the Run 1 \cite{Abi2021} and Run 2/3 result \cite{PhysRevLett2023} as well as the FNAL average of the two. The Brookhaven result \cite{Bennett2006} is also shown and the combination of all the above is shown as the world average \cite{PhysRevLett2023}. \textbf{Right}, Error budget of the Run 1 analysis compared to the Run 2/3 analysis. }
    \label{fig:syserrandresults}
\end{figure}

As shown in Figure \ref{fig:syserrandresults} all systematic uncertainties were reduced in Run 2/3. Overall, the systematics were improved by more than a factor of two. The greatest improvement came from the replacement of damaged high voltage resisters in the ESQ plates.  These plates which pulse in synchronisation with the muon bunches, are individually regulated by 32 resistors. This issue introduced an asymmetric time dependence in vertical focusing, which in turn affected the stability of the radial and vertical position and width of the beam. Fixing these resistors thus drastically decreased the uncertainty on the phase acceptance corrections, $C_p$. 

The oscillating magnetic fields from vibrating quads were measured with a new and improved NMR probe. Measurements were also taken at many more positions around the ring; this, combined with repeated measurements over time, reduced the transient magnetic field correction $B_q$ uncertainty significantly. An improved magnetometer also reduced the systematic error of the kicker transient $B_k$ by a factor of $\sim 3$. 

The kicker system is a non-ferric magnet that "kicks" the injected muon bunch into a stable orbit around the storage ring. During Run 1 and Run 2, the kicker voltage was limited to 142 kV due to cable constraints, preventing the kick from perfectly centring the muon beam and storing the muons at the so-called "magic momentum", which minimises the electric field correction. During Run 3 the cables were upgraded, enabling the kicker to operate at a higher voltage. This improvement led to a muon beam that was better centred, reducing the electric field correction $C_e$ and decreasing the amplitude of Coherent Betatron Oscillations (CBO), a beam frequency that arises from the radial motion of the muon beam. 

Systematic uncertainties in the measurement of the precession frequency were also reduced. The sources of these uncertainties arise from effects that cause a frequency bias. In Run 1 the systematics for $\omega_a$ were dominated by the correction of pileup contamination and modelling of the CBO.  A significant effort was applied in Run 2/3 to reduce both of these systematic uncertainties, among others. 

Pileup arises when the reconstruction algorithm is unable to distinguish between two closely timed positrons hitting a calorimeter. An improved clustering technique (crystal hits are grouped together to reconstruct a positron), which used the energy dependence of the timing resolution, reduced the pileup contamination and thus the uncertainty from 35 ppb in Run 1 to 7 ppb in Run 2/3 \cite{Foster2024}.

The main challenge of modelling the CBO lies in accurately modelling the decoherence of the CBO signal and the time dependence of the CBO frequency. Despite remaining the main source of uncertainty in Run 2/3, the increased statistics enabled the testing of more refined models to tackle this challenge. The increased statistics together with the improved stability of running conditions have reduced the uncertainty from 38 ppb in Run 1 to 21 ppb in Run 2/3.

The Muon g-2 Collaboration announced its second measurement of $a_{\mu}$ on  August 10 2023 with a precision of 215 ppb a factor 2.2 lower than Run 1 \cite{PhysRevLett2023}. The Run 2/3 uncertainty contains a systematic uncertainty of 70 ppb, which surpasses the experiment’s 100 ppb design goal \cite{grange2018}. The Run 2/3 result is in excellent agreement with Run 1 and the Brookhaven result. Both FNAL results,  the Brookhaven result, and the combined world average are shown in Figure \ref{fig:syserrandresults}. The experimental world average (the Run 1 + Run 2/3 FNAL and BNL measurements) at the time of writing is $a_\mu (\text{exp})= 116 592 059(22)×10^{-11}$.

\section{A comparison with the Standard Model prediction}

The g-factor is sensitive to all sectors of the SM making the calculation of the muon $g-2$ highly complex, involving contributions from QED, electroweak, and hadronic effects. Hadronic effects arise from vacuum fluctuations in strongly interacting particles, known as hadronic vacuum polarization (HVP). These effects are most difficult to calculate and thus provide the dominant uncertainty in the calculation of $a_\mu$.

\begin{figure}[htbp]
	\centering
	\begin{subfigure}{0.45\linewidth}
		\includegraphics[width=\linewidth]{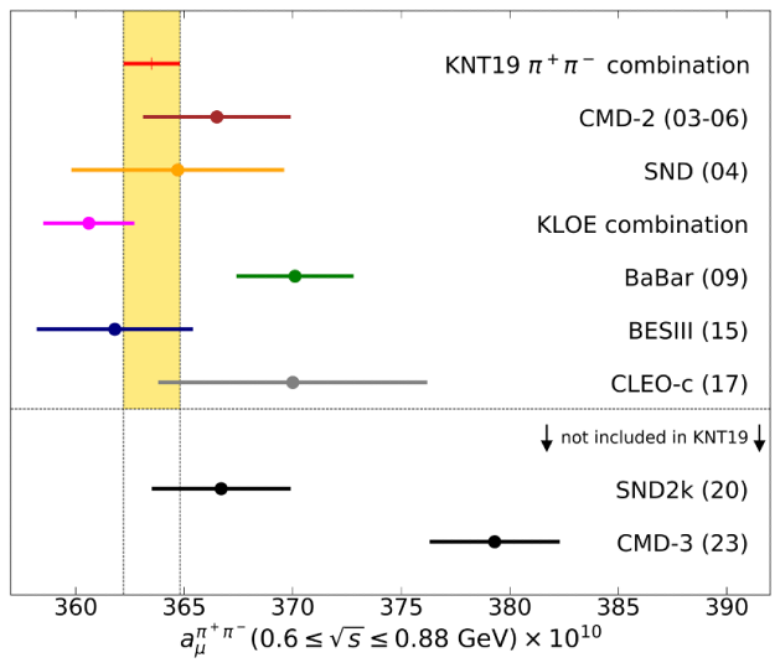}
        \label{fig:run1res}
	\end{subfigure}
	\begin{subfigure}{0.53\linewidth}
		\includegraphics[width=\linewidth]{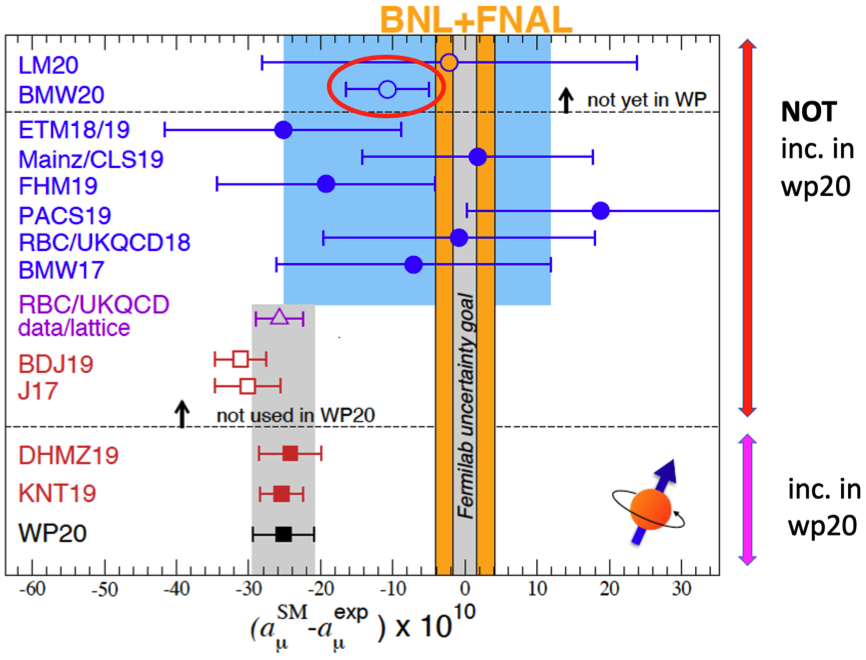}
        \label{fig:syserr}
	\end{subfigure}
    \caption{\textbf{Left:} The dominant $2\pi$ contribution to the HVP provided by different $e^+e^-$ experiments \cite{Venanzoni2023}.   \textbf{Right:} Summary of the muon $g-2$ landscape comparing theoretical predictions with the experimental measurement of $a_{\mu}$ \cite{Colangelo2022}. }
    \label{fig:theoryresults}
\end{figure}

In 2020, the Muon $g-2$ Theory Initiative updated the Standard Model (SM) prediction \cite{Aoyama2020} with 0.37 ppm accuracy using the dispersive approach, which relies on $e^+e^-$ to hadrons cross-section measurements. This value was further confirmation of the persistent gap between theory and experiment and provided intriguing hints of new physics. However, in 2021, the BMW collaboration presented a Lattice-QCD calculation of HVP that was closer to the experimental value and thus in $2.1 \sigma$ tension with the dispersive approach. Further complications arose in 2023 when the CMD-3 experiment measured the $ e^+e^- \rightarrow \pi^+\pi^- $ cross-section \cite{Ignatov2024}, reducing the discrepancy between the experimental value and disagreeing with previous dispersive results. The current landscape of the g-2 puzzle is even more interesting in light of new theoretical predictions from lattice QCD in 2024 \cite{Boccaletti2024} \cite{Kuberski2024}.  These ongoing tensions in the hadronic sector continue to challenge a definitive comparison between theory and experiment for the muon $g-2$ measurement.

\section{Conclusion and outlook}

The Muon g-2 Collaboration has achieved an unprecedented precision of 215 ppb in measuring the muon magnetic anomaly, more than halving the uncertainty of its initial 2021 result. This new measurement aligns with both the earlier result and the Brookhaven experiment. Looking ahead, the collaboration will release a new and final result in 2025, analysing three additional years of data, exceeding its initial goal of collecting 21 times the Brookhaven dataset. This, together with  improvements in running conditions means the collaboration is on track to further refine and enhance its precision.

However, recent developments in Standard Model theory have made comparisons between experiment and theory more challenging. The Theory Initiative will be producing a new SM prediction for $a_\mu$ in 2025 that will better reflect this landscape. Additionally, of the groups shown in figure \ref{fig:theoryresults} which provided a dispersive result, many have ongoing analyses which could provide a clearer answer in the upcoming years. One such avenue is the MuonE experiment \cite{Pilato2022},  which  aims to precisely measure the hadronic vacuum polarization (HVP) contribution to the muon $g-2$ by analysing elastic muon-electron scattering at the CERN SPS, providing an independent, data-driven determination to resolve discrepancies between theoretical predictions and experimental results.

\section{Acknowledgments}

This work was supported in part by the US DOE, Fermilab, the Science
and Technology Facilities Council (UK), the Royal Society (UK), the European Union Horizon 2020 research and innovation program under the Marie Sk\l{}odowska-Curie grant agreement No. 101006726, and the Leverhulme Trust, \texttt{LIP-2021-01}.

\end{document}